\documentclass[journal]{IEEEtran}
\ifCLASSINFOpdf
  \usepackage[pdftex]{graphicx}
\else
  \usepackage[dvips]{graphicx}
\fi
\begin{document}
%
\title{A Community Membership Life Cycle Model}

%
%
%
\author{
\IEEEauthorblockN{Andreas C. Sonnenbichler}
\IEEEauthorblockA{
Information Services and Electronic Markets\\
Institute of Information Systems and Management\\
KIT, D-76128 Karlsruhe, Germany\\
Email: andreas.sonnenbichler@kit.edu}
}

\maketitle

\begin{abstract}
Web 2.0 is transforming the internet: Information consumers become information producers and consumers at the same time. In virtual places like Facebook, Youtube, discussion boards and weblogs diversificated topics, groups and issues are propagated and discussed. Today an internet user is a member of lots of communities at different virtual places. “Real life” group membership and group behavior has been analyzed in science intensively in the last decades. Most interestingly, to our knowledge, user roles and behavior have not been adapted to the modern internet. In this work, we give a short overview of traditional community roles. We adapt those models and apply them to virtual online communities. We suggest a community membership life cycle model describing roles a user can take during his membership in a community. Our model is systematic and generic; it can be adapted to concrete communities in the web. The knowledge of a community's life cycle allows influencing the group structure: Stage transitions can be supported or harmed, e.g. to strengthen the binding of a user to a site and keep communities alive.
\end{abstract}


\begin{IEEEkeywords}
Community Membership Life Cycle Model, Virtual Communities, Online Communities, Life Cycle, Social Network Analysis
\end{IEEEkeywords}

\section{Introduction}
\label{sec:introduction}
Web 2.0 is ubiquitous in the net. Personalization, customizing, user-created content and participation of the many are the fundament, Web 2.0 is built on. In the old internet, virtual places ("sites") communicated mainly unidirectional: A fixed number of information producers (authors) created the content for masses of information consumers, the "site visitors". Web 2.0 is changing this by making everybody not only an information consumer but also an information producer. Toffler (\cite{toffler_third_1984}) suggested the term "prosumer". A new kind of mass communication media is currently being established, letting masses of information creators communicate with masses of information consumers. 
Thousands of different virtual places have been built: Discussion boards on every thinkable topic are available in the net, where one can discuss the latest news of roadsters, the Swine Flue or politics. Weblogs and Twitter \cite{twitter} are changing the media and the journalistic world enabling us to participate, for example, in Iran's current political development. Facebook \cite{facebook} as a social networking platform lets users share personal content and media. Flickr \cite{flickr} is specialized on pictures. Xing \cite{xing} describes itself as a business platform. All this places we call virtual communities. 
The phrase "virtual community" has first been used by Rheingold \cite{rheingold_slice_1994, rheingold_virtual_2000}. He was participating in an early online community called the WELL (Whole Earth 'Lectronic Link). Already in 1994 he wrote "... virtual communities are cultural aggregations that emerge when enough people bump into each other often enough in cyberspace. A virtual community is a group of people who may or may not meet one another face to face, and who exchange words and ideas through the mediation of computer [...] networks" (\cite[p.57]{rheingold_slice_1994}).
Rheingold describes virtual online communities similar to real life groups and communities: "In cyberspace, we chat and argue, engage in intellectual discourse, perform acts of commerce, exchange knowledge, share emotional support, make plans, brainstorm, gossip, feud, fall in love, find friends and lose them, play games and meta-games, flirt ... We do everything people do, when people get together, but we do it with words on computer screens" \cite[p.58]{rheingold_slice_1994}. 

Complementary, Lazar and Preece \cite{lazar_classification_1998} defined four attributes of a virtual online community: 
\begin{enumerate}
\item People, who interact socially to satisfy their needs and/or perform roles. 
\item A shared purpose, such as an interest, need or service as a common goal. 
\item Policies in the form of tacit assumptions, rituals, rules or guides. 
\item Computer systems, to support the social interaction.
\end{enumerate}

We argue, along Rheingold and Lazar and Preece, that virtual communities can be seen in some ways like traditional, offline communities. 

In chapter \ref{sec:motivation} we will give examples, how a systematic, generic community membership life cycle can create benefits. Chapter \ref{sec:relatedwork} gives an overview of related work. Our own approach to a community membership life cycle will be introduced in section \ref{sec:a_community_membership_life_cycle_model}. We finish this article in chapter \ref{sec:conclusion} with a conclusion and references to future work.

\section{Motivation}
\label{sec:motivation}
Success of virtual communities is closely related to the group structure of its members. We expect many sites, like Facebook, Youtube \cite{youtube}, Xing, Flickr, the local chess club, a MMORPG\footnote{Massively Multi-Player Online Role Play Game. These are games played online with several hundreds or thousand players.} 
guild and many more to be interested in as many members as possible. Many users make a platform a quasi-standard in its competence field. By this, depending on the individual business model, we expect advertisement fees, direct user fees or other forms of income to increase. The importance of a critical mass of users is pointed out for example by \cite{armstrong_real_2000}.
Increasing the number of community members means on the one side to acquire new users and on the other side to motivate members to stay. 

The same argument applies not only to professional, commercial sites, but as well to private discussion boards on specialized topics. Of course, the target groups of sites differ: Facebook wants to attract different members than a private discussion board of the regional computer club. However, every internet site can be expected to be interested in gaining new members within their target group and binding existing members to the community.

One of the major questions therefore is how to commit users to the community. One way to achieve this is, to define precisely the target groups a virtual community is made for. This strategic decision influences the content, layout, personalization, marketing, customer communication and many more. Further more, the inner structure of a virtual community should be known, to be able to influence its development:
\begin{itemize}
\item In which stages of a community membership life cycle can a member be? 
\item How can we identify these stages?
\item What distribution of the users on these stages can be observed?
\item What stage transitions are likely to happen?
\item How can we influence transitions and how can we gain the target distribution?
\end{itemize}
This inner structure of a virtual community is what we will focus on in our article. 

Let us assume a virtual community of a local chess club with several hundred members. The pure number of users seems quite high. We argue, that the number alone does not tell much about the community: Do the members participate in discussions? If so, what actions have to be taken to stabilize the community? If not, how can the community become more active? How many members only signed up and never came back again? How can they be motivated not only to sign up but stay? Who are the informal community leaders? Shall they be kept at this number, shall there be more or less of the leaders?
We see the steering of the inner structure of a community as an important step in keeping a community healthy. The community owner shall be given an analytic instrument to be able to compare his community with and from that point take actions to steer his community.

Furthermore, the knowledge of the inner structure of a community combined with automatic detection of a user's current community role can be used to personalize the user's community environment. A user, who is automatically identified as a new member, can be offered a guided tour and explanations of community rules. An experienced user can be offered special features like moderating discussion boards, sort, and administrate picture galleries and video streams. In case a user is identified as a troublemaker he can be limited in his actions. 

To be able to offer such automatic community services, two preconditions must be met: The inner structure of a community with its roles must be known. And measures for automatic identification of users having these roles must be defined.

\section{Related Work}
\label{sec:relatedwork}
In this section we will refer to observations and models dealing with the development of communities and groups offline and online. 

\subsection*{Small Group Dynamics}
\begin{figure*}[ht]
	\begin{center}
		\includegraphics[width=0.8\textwidth]{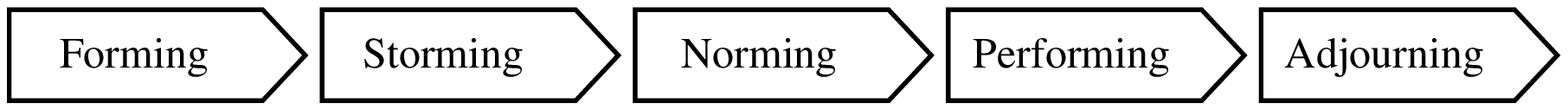}
	\end{center}
	\caption{Small Group Dynamics suggested by \cite{tuckman_stages_1977}}
	\label{fig:tuckman} 
\end{figure*}

Tuckman and Jensen \cite{tuckman_developmental_1964} did research about stages in group development. He suggested the four well-known stages forming, storming, norming and performing. Later in 1977 Tuckman and Jensen \cite{tuckman_stages_1977} presented a fifth stage called adjourning. The phases are depicted in figure \ref{fig:tuckman}.

In the forming phase, the group members orientate themselves and test the reactions of others. The storming phase is full of hidden or open conflicts partly with resistance in the group. This phase is followed by the norming stage, where each group member finds his place and norms for behavior are determined and agreed on. After this a constructive performing phase follows, succeeded by the adjourning stage with anxiousness about leaving the group and feelings toward leaders and group members. Tuckman did his research for this model on small groups and on group dynamics. The analyzed groups consisted of roughly a dozen members. Of course, in the '60s and '70s virtual communities were not analyzed. 

\subsection*{A Remote Masters Program}
The development of a combined real world and virtual community is described by Caroline Haythornthwaite \cite{haythornthwaite_community_2000}. A remote masters program offered by the University of Illinois consists of a boot camp, where students physically meet on the campus at the beginning of the program. From then on, the classes meet only virtually, coming together physically only once per year for a day. The virtual community formed by each cohort used tools like PowerPoint for lectures, Internet Relay Chat (IRC) for questions and web boards for discussions and exercises. 

Haythornthwaite describes the initial phase at the boot camp as "initial bonding phase". Here, in a traditional way, contacts are established, group processes take place. After this initial come together a "maintaining presence phase" is observed. In this phase, "maintaining ties and community at a distance [...] is perceived by students to require more effort than in a face-to-face community" \cite[p.17]{haythornthwaite_community_2000}. The third phase described is the "disengaging from the community". As the community members "progress through the program, the desperate need to make contact diminishes. They become familiar with [...] routines, the technologies and norms for their use, and their distanced companions and fellow travellers" \cite[p.23]{haythornthwaite_community_2000}. Other terms for the three phases can be found, Johnson \cite{johnson_survey_2001} calls these phases initial bonding, early membership and late membership. We will refer to the initial notion used by Haythornthwaite.

This research was done on medium group sizes (about 30 to 50 members), in comparison to Tuckman, who did research on smaller groups. Haythornthwaite's model represents medium sized groups with a common target, namely to receive the masters degree.

\subsection*{Communities As Products}
Owyang \cite{owyang_online_2008} did research on successful commercial online communities describing them as commercial products. So he "interviewed 17 people ("many community leaders that you know") to find out the commonalities between successful communities". He determined a 7-phases life process a successful community passes through. 

\begin{figure*}[ht]
	\begin{center}
		\includegraphics[width=0.8\textwidth]{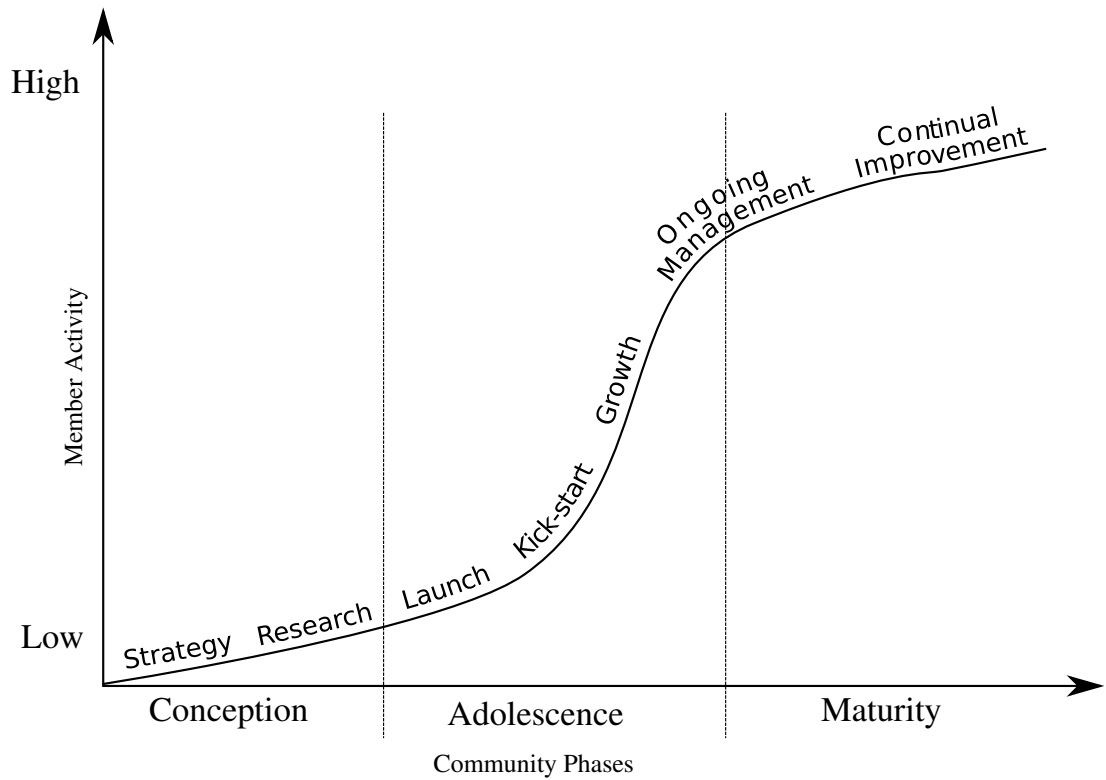}
	\end{center}
	\caption{Life process of a successful virtual community by \cite{owyang_online_2008}}
	\label{fig:owyang} 
\end{figure*}

As depicted in figure \ref{fig:owyang}, three major phases are distinguished: conception, adolescence and maturity. In the conception phase the stages strategy and research are differentiated analyzing market situation, target for the community, and usage for the stakeholders. During launch and kick-start, as stages of the adolescence phase, the successful community gains most of its users which is critical for its success. At the end of this phase and also at the beginning of the maturity phase, ongoing management and continual improvements need to be done to keep the community successful. Member activity mapped on the y-axis is a relative measure, not further explained by Owyang. 
His research is mainly on large communities, which potentially have hundreds of members. A main focus of his work is about, how to steer processes and instruments necessary to gain a critical mass of community members, to create a commercially successful community.

\subsection*{Building Online Communities}
\begin{figure*}[ht]
	\begin{center}
		\includegraphics[width=0.8\textwidth]{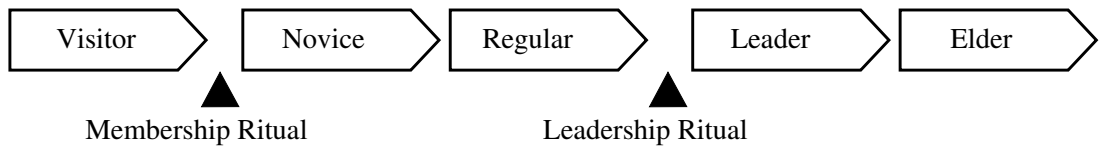}
	\end{center}
	\caption{Community Life Cycle by \cite{kim_community_2000}}
	\label{fig:kim} 
\end{figure*}

A community-centered life cycle is described by Kim \cite{kim_community_2000}. She distinguishes five stages in three main life cycle steps (see figure \ref{fig:kim}).  

Visitors are new, have not signed up for an account and do not have an identity in the community (yet). They are unfamiliar with local customs, techniques, behavior and have many unanswered questions. 
The next step in the life cycle are novices. They have signed up for an account, but still have to learn the ropes and be introduced to the community life. "Novices need to learn what they can do, whom they can do it with, where they can do with, and how they're expected to behave" (\cite[p.133]{kim_community_2000}).
Becoming more experienced in the community, the member becomes a regular. Regulars are established members, the mainstays of a community. Regulars know the environment and opportunities, know, how to find what they are looking for, how to personalize their interface and how to communicate efficiently with other members of the community. 
The fourth step are leaders. They are the ones, who "help newcomers get settled in, operate the community shops and taverns, volunteer for charities and committees, and run for mayor" (\cite[p.119]{kim_community_2000}). They answer questions and help members to solve problems with the system. Leaders plan, coordinate and run events in the community and might provide special resources or services to members. 
The last step in the life cycle are the elders. "Over time, some leaders will tire of their day-to-day activities and step down from their official roles. Because they're familiar with the history and inner workings of the community, they're now elders - respected sources of cultural knowledge and insider lore. Along with other long-time residents, they're the teachers and storytellers of the community, the people who give the place a sense of history, depth and soul" (\cite[p.119]{kim_community_2000}). 

Kim gives us with the 5-stage life cycle a reasonable model to distinguish different user roles in already established communities of large size. 

\subsection*{Online Learning Communities}
Another description of roles in virtual communities is given by Palloff and Pratt \cite{palloff_building_1999}. They investigate inner mechanisms in online learning communities. Especially in discussions about new lectures' content three roles for the students unfold: Knowledge Generators, Collaborators and Process Managers. 

Knowledge Generators are people, who actively assimilate knowledge by constructing new forms of knowledge or meaning. They combine different knowledge together and present new results through these combinations. We see this role is not limited to learning communities: People, who know about other facts currently discussed or available, bring in new sources, new ideas and knowledge in the community. Therefore Knowledge Generators are not only a role in learning communities, but a very general role. 

Collaborators assist the community in making sure, that all voices are heard and all members are participating. They will allow a group not to forward until a consensus has been achieved and might use for this tools like web surveys and ratings. On the other hand collaborators do not work only within a community but also between them: So they connect two or more groups helping all of them through information and knowledge exchange. Again we see collaborators not limited to learning communities: Every community might have or need people caring about that nobody is left behind or mediating knowledge between different groups. 

Process Managers are helping to maintain the process, slowing down discussions or progress if they feel, they or somebody else is lost, holding the direction, if discussions tend to get of the path, feeling generally responsible for the group moving in the right direction. 

Palloff and Pratt describe three roles in learning communities that can easily be adapted to general communities. Knowledge Generators, Collaborators and Process Managers can be found probably in any virtual community: People, who create and have new ideas, and make new suggestions; Collaborators, who take care about the group processes; Process Managers or moderators acting as facilitators to help a community to reach an explicit or implicit goal. In contrast to other models not the development of the group or the group processes are in the focus, but different group roles. 

\subsection*{Summary}
We have described several life cycle models for (virtual online) communities. Reviewing these models closely we notice that three different perceptions of the term life cycle exist: 
The first meaning focuses on the development of a single community member and the roles he takes over time. This applies to the model of Kim. We call this perception \textit{community membership life cycle}. The second understanding focuses on group processes, a community undergoes as a whole over time. Here not so much the individual development is in the spotlight, but the inner group processes. This we will designate as \textit{group life cycle}. Examples for this perception are given by Tuckman and Jensen. The third interpretation sees the community as a product, evolving over time. This we define as \textit{community product life cycle}. Owyang is a representative for this approach.

\begin{figure*}[ht]
	\begin{center}
		\includegraphics[width=0.8\textwidth]{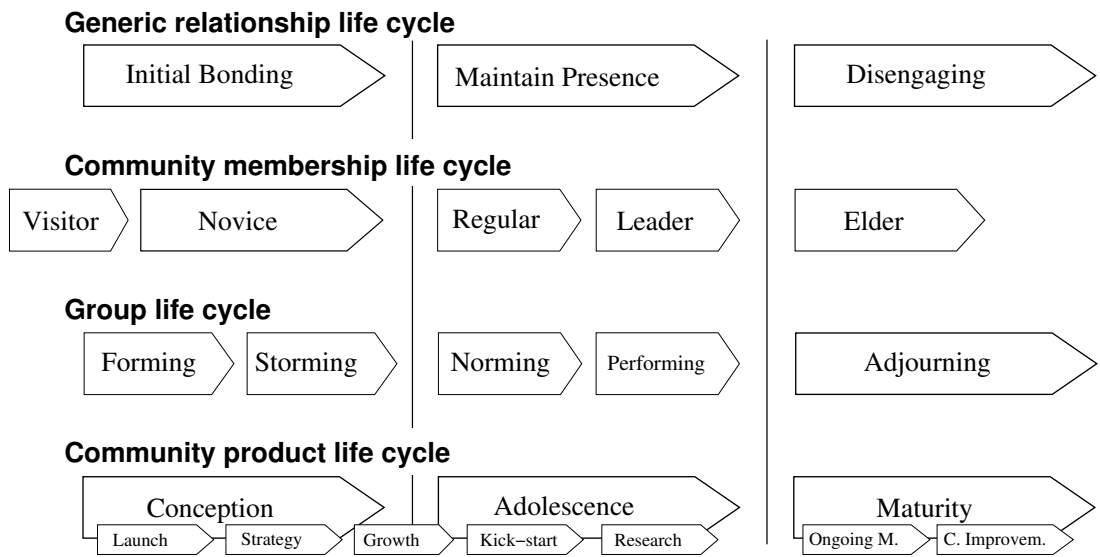}
	\end{center}
	\caption{Comparison of life cycle models}
	\label{fig:comparison} 
\end{figure*}

A comparison of the life cycles can be found in figure \ref{fig:comparison}. As depicted a generic model is given by Haythornthwaite. In her model, three main phases “initial bonding”, “maintaining presence” and “disengaging” are described. We argue that this model can be seen as a generic one, as all other models can be seen as finer-grained specialisations.

Kim's life cycle model starts with an earlier phase, which can not be found studying the other models: The first role here is a visitor, who is not (yet) part of the community. He has not signed up and decided whether he wants to become a part of the community. Kim's novice stage corresponds in some ways with Haythornthwaite's initial bonding phase. In the proximate phase, Kim differentiates between regulars and leaders, both fully partaking roles in the community, but at distinct activity levels. Leaders are fully committed to the community, identify themselves a lot with it and try to push the community forward very actively. Partly, also the “elder” belong to Haythornthwaite's “maintaining presence” phase, as Kim describes them quite participating and active in communities as storytellers and the soul of a community. On the other hand, the relation of the “elders” to the community diminishes, linked with Tuckman's adjourning phase.

Tuckman and Jensen's model is finer grained than Haythornthwaite's generic model. In comparison to Haythornthwaite's initial bonding phase, Tuckman and Jensen split this phase into forming and storming. “The need to present oneself to the group and define its own place”, how Haythornthwaite describes her first phase corresponds with Tuckman's initial stages. We see the norming and performing phase linked the maintaining presence of Haythornthwaite. The members found their place within the community (or find it, in the norming phase) and start to be a productive part of the group. In both models, the last phases correlate almost perfectly, describing the same situation in a community, namely the disengagement, the dissociation of one or more individuals of the group. 

Owyang describes a different perspective, describing a community as a commercial product. Although he does not focus on inner group processes, his model fits well into the generic model of Haythornthwaite. Again, we see a three part model with the three main phases conception, adolescence and maturity. These main phases are finer structured and described by sub-phases. Of course, Owyang describes the life cycle of a whole community and not of its members.

Not represented in the figure is Palloff and Pratt's model. They observe three different roles of students in learning communities: Knowledge Generators, Collaborators and Process Managers. As these roles are relatively steady over time and explain more the behavior of a student and how they contribute to the group, there is of course no sense in interpreting these findings as a life cycle model. But these roles can be used to characterize different roles in the performing phase of a group life cycle.

\section{A Community Membership Life Cycle Model}
\label{sec:a_community_membership_life_cycle_model}
Building or maintaining a modern virtual community means to get and keep members. A structured community membership life cycle model can help to define an inner community target structure and to analyze and compare this with the current inner structure. Actions can be defined to transform the current structure to the target structure. A structured community membership life cycle can provide the fundamentals this analysis and actions are based on. 

In this chapter we suggest a generic community membership life cycle model. This model can be applied to concrete communities. 
For each role we describe, how it can be identified, how a typical personal relationship to other members looks like, what interests and needs the member has and what successor roles are taken usually. 
To identify roles, we will suggest measurements from the Social Network Analysis (SNA) and activity measures. Typical SNA measures are degree, closeness, betweenness and Eigenvector centrality. For a good introduction we refer to \cite{wasserman_social_1994}. 

Besides SNA we suggest activity measures for role identification. Examples are the time since the last login or clicks/operations performed during a sliding time period. All absolute measures will have to be measured relatively regarding the general “level” of the community: In a very active community, for example, five discussion contributions can be very low compared to some opinion leaders, posting 50 posts a day. In a less active community, five contributions can indicate a leadership position. 
We see our model as generic. If this model is applied to a concrete community in the web, it is necessary to test for the fitness of the roles: Is it necessary to split a generic role into sub roles? Is a role applicable?

\begin{figure*}[ht]
	\begin{center}
		\includegraphics[width=0.8\textwidth]{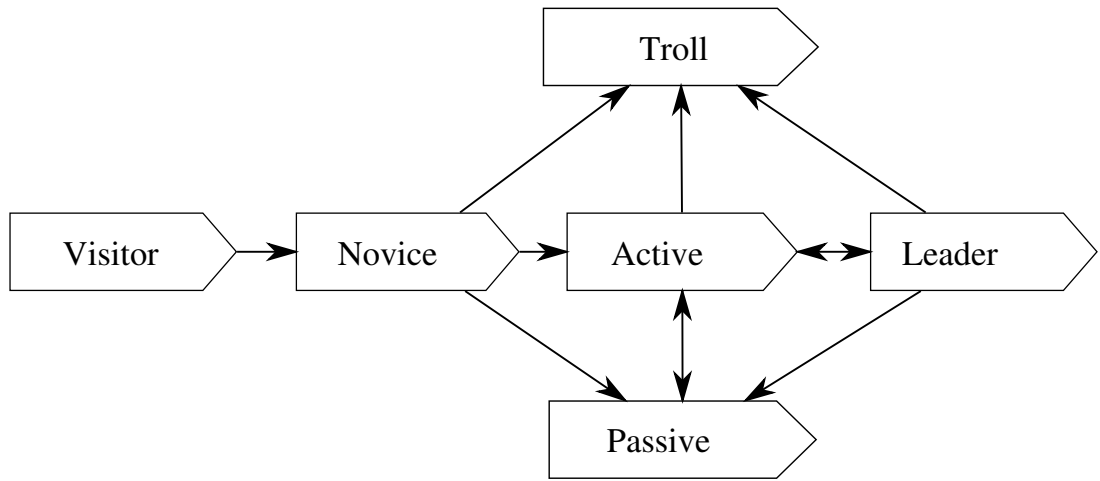}
	\end{center}
	\caption{A Community Membership Life Cycle Model}
	\label{fig:community_membership_life_cycle} 
\end{figure*}

\begin{figure*}[ht]
	\begin{center}
		\includegraphics[width=0.8\textwidth]{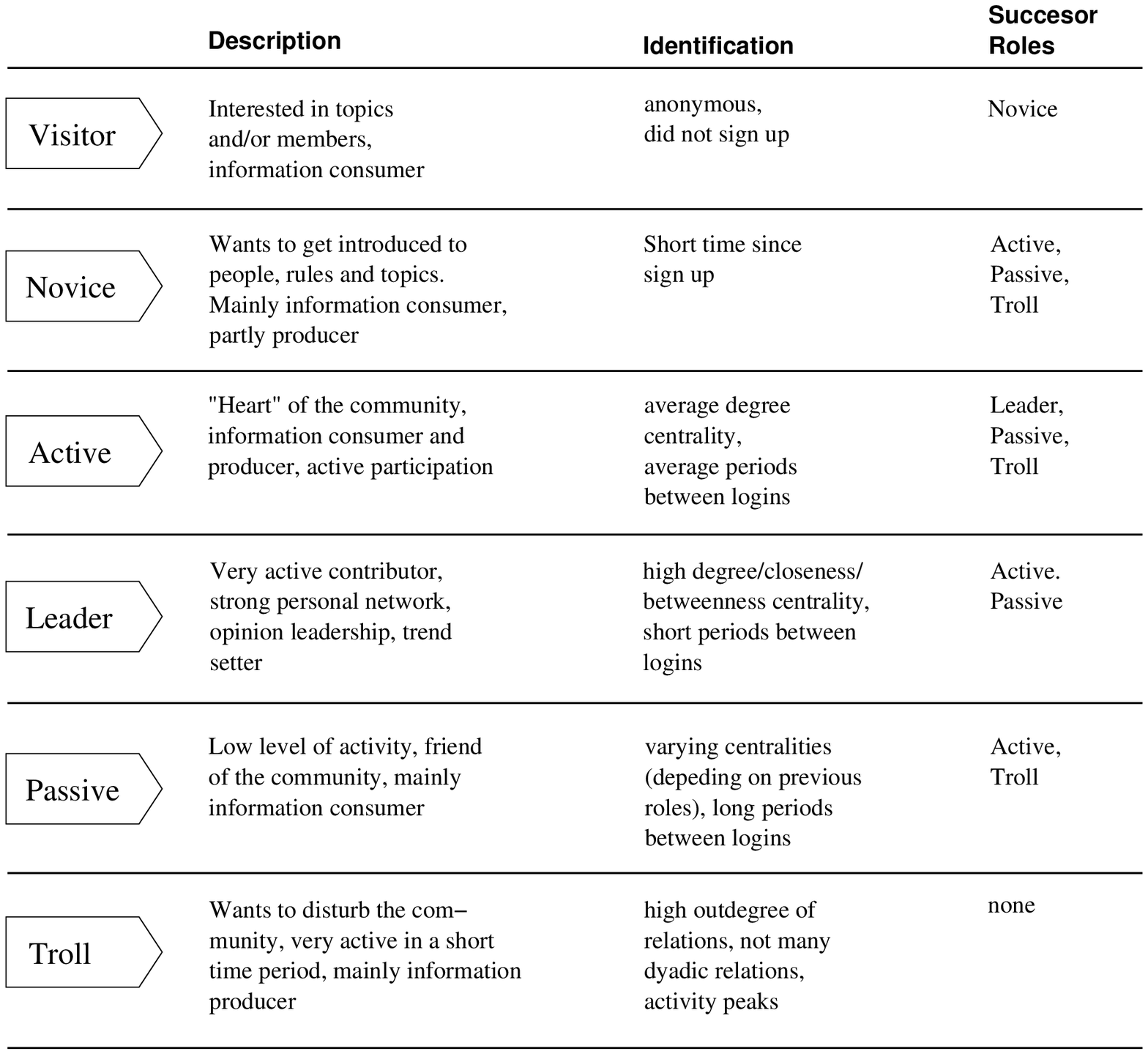}
	\end{center}
	\caption{Overview of Community Membership Life Cycle Roles}
	\label{fig:community_membership_life_cycle_roles} 
\end{figure*}

The generic community membership life cycle model is depicted in figure \ref{fig:community_membership_life_cycle}. The initial role for a new user is the role of a visitor. A description of all roles will be given below. The user can become a novice next. The successor roles of a novice are passive members, active members or trolls. Actives can become passive, leaders or trolls. We expect Leaders to down step to actives, passives or become trolls. Passive members may become active again.
The roles of Palloff and Pratt are missing; they can be used as sub roles.
An overview on the roles is given in figure \ref{fig:community_membership_life_cycle_roles}.

\subsection*{Visitors}
The first contact a new member has with a community takes place as a visitor. He has not signed up, has not got an account and, therefore, has no identification and no name in the community. He does not know much about the community, maybe a little about the topics or some members, depending on how he learned about the community. If somebody recommended the community to him, it is likely, that he knows at least one community member. In case he found the community by searching for a topic, he probably knows about some discussions in the community. 
We expect him to be interested in finding out more about the community: Who are the members and what are the topics? Is this community interesting enough for him to join?
After gathering some information about the site, he has to make a decision: Sign up and become a member or leave. Depending on the structure of the community, he can even decide to stay as an anonymous visitor, come back from time to time and read articles, watch videos or pictures. E.g. many blogs might have a huge number of visitors which stay in this role. In contrast, closed discussion forums like guild forums for online games may allow visitors only to sign up and not read any content without this. In such a case, the visitor phase is very short.
Identifying a visitor can be realized with many technologies: Click streams analysis via IP addresses, cookies, hidden arguments, session ids or logs integrated in the application. 

\subsection*{Novices}
As soon as somebody signed up he becomes a novice. Now he can be identified within the community by his nickname. He still does not know much about the community. He is willing to get introduced to people and/or issues. He may want to build up a personal network with other community members. We expect him to be willing to learn about explicit and implicit rules, behavior and important people. 

A novice can become an active member or a passive member (see later) or leave. We argue that the role of a novice is a temporary one. The first orientation phase is over quite soon. Then the decision has to be made, to participate actively, become passive or leave.

A novice can be identified reliably by the time since he signed up. Usually the date of his sign up is stored in the user profile and can be compared with the current date. 

\subsection*{Actives}
Actives are the backbone of a virtual community. They participate actively in the community life, both as consumers and producers (prosumer). They read content like posts, pictures, videos and contribute with own content. They make contacts to other users and build up a social network. They are the pulsating heart of a community. We expect that the interests of an active are to participate in the community life, consume existing content, find new content and contribute. Furthermore, we expect an active to foster existing social relationships and gain new contacts. An active is interested in showing up regularly to inform him about the latest news.

We see many successor roles of an active. Of course, the member can continue being an active. In contrast to a novice, an active is not necessarily a short-term role. An active can decide to become passive and participate less in the community. He can become a leader in the community. Or, on the opposite, quit immediately. We do not expect an active to quit immediately directly, but expect him first to become passive and then quit. On the other hand, we can think of situations in a community, were an active quits from one day to the other. E.g. a clash in the community might take place or events outside of the community may happen.

Actives can be identified through a combination of measures. First, as the name says, they should be active. So the time since the last login should be average or low compared to all community members. The average time between the last logins should be also average. On the social side, we expect an active to have an average degree centrality. 

\subsection*{Leaders}
Leaders are people, who run communities, plan events, moderate discussions and may even administrate communities technically. They contribute very actively in content, discussions and media. Due to their activity level, they have strong ego-networks which they maintain and increase. They are the community experts, who know members, content and techniques. Leaders have many and close relationships. Their interests can be described with a high commitment with the community. They want it to grow, be alive and have many active members. As leaders they may be interested in their role and their special position.

If the interest in the community diminishes, a leader can step back and become passive. Some may also choose to become regular actives. And of course, a leader can also quit; although, we seldom expect leaders to quit immediately.

We expect a leader to have a high degree centrality. As leaders of opinions we expect them to have high betweenness, closeness or Eigenvector centrality. A leader is characterized as having an activity above a certain threshold. By this, we expect the average time between logins not to be lower than that of actives.

\subsection*{Passives}
A role with a lower activity level than leaders or actives are the passive members. Many offline communities, like a tennis club, have not only active members or leaders, but passive members. Actually, in many communities the number of passives can be much higher than those of the actives. They are the “silent supporters” who like the community topics and/or members and want to stay in a loose contact. They are interested in news and the people but do not want to participate in an active way. They normally are rather consumers than producers. We expect them to show up not very regularly or in larger time frames than actives.

Successor roles of passives are the actives: If a passive is motivated in some way, he can become active (again). Passives are also likely to quit, if their interest in the community diminishes. We do not expect that a passive becomes a leader immediately, but - at least for a short time - becomes an active first.

To identify a passive we expect him to have an average time between logins which is lower than the average of all members. Being more a consumer than a producer of content we can identify a passive as well by a low number of posts. On the social network side, we expect passives to have a very stable social network over time. Not many new contacts will be made and not many contacts will be lost. A passive can have a wide range of networks, from a very small personal network, in case the user became passive immediately after he joined the community, to a very large network, if he was a leader beforehand.

\subsection*{Trolls}
An additional role we want to introduce, we call Trolls. The role describes a negative, disturbing trouble maker in a community. The word refers to mythological trolls, which are said to be crabby and bad-tempered. We chose to use this term, as it is quite common in network culture (cf. \cite{dhala_simple_1999, raymond_new_2009}). Trolls are users, who at least disturb other members by posting offending, improper content like flame messages, forum posts, pictures or other media. Spamming is a typical troll activity. We introduce this new role as in contrast to “real life” in a virtual community it is easier to behave in a Troll's way. In many cases, members are anonymous in the web. Users can create multiple user accounts. To abuse one account for troll activity leads to less or no consequences for the real user, as he can act anonymously. If a user account is identified as a troll and lost its reputation, a new account can be created. If effective mechanisms exist in a community, to avoid multiple accounts (e.g. by identifying a user by his real life name), we expect much less trolls to occur.

Trolls do not have successor roles: As the reputation of Trolls is normally very low, we expect it to be very hard or impossible for them, to return to a regular role like an actives or a passive. We see it more likely, that a user with the role troll might sign up again with a different account and start a new membership life cycle.

Trolls are expected to have either not many connections to other community members, as their account is usually made just for this negative purpose. Or they have a very high out degree of relations, which are not accepted by others. The activity level of trolls is usually very high in a short time period and reduced to a very low level. We expect a troll trying to cause as much furore as possible and disappear afterwards without a trace. We see a troll's interest in disturbing, provoking or abusing the community.

\section{Conclusion and Future Work}
\label{sec:conclusion}
The knowledge about a community membership life cycle presents a model for the inner structure of a virtual community. We suggested six different roles for this cycle. Visitors are interested in the community and information consumers. Novices sign up and are newbies who want to get introduced into the community, its topics, members and rules. Actives are the heart of a community, the ones who participate in the community life in an active way. They are information consumers and producers at the same time. Leaders are the “rule-makers”, the opinion makers and the one's who are highly committed to the community. Leaders are trend-setters concerning information production. Passives are members still interested in the community topics and/or its members, but rather consumers than information producers. Trolls are the negative side, the ones who want to disturb and cause trouble.

This generic model defines life cycle stages of community members. We presented measurements, how the roles can be identified e.g. by typical SNA measures. For a concrete virtual community, the model allows to observe the distribution of members over the roles. This knowledge can be used to influence this distribution. A community with many passive members might want to reactivate passive members, so that more new content is produced and the community is more active again. A community with many visitors may want to provide mechanisms to win visitors as new members. 
We see the benefit of the life cycle model two-fold: It makes distributions of roles measurable. And it can help to define actions to bring the community structure to a target distribution. The first helps, to identify an inner structure of a community and analyze its distribution. This provides in-depth knowledge about the community. The latter enables mechanisms to manage communities and influence its inner structure.

In future work we plan to analyze showcases. We see the following steps necessary to apply our generic model to a concrete community. First, check, whether more sub roles need to be defined or a role can be omitted. E.g. for an a-priori closed community, the role visitor is unnecessary. Second, define thresholds for all measurements for identifying the roles. So we can define an active having a 20\% lower average time between logins than the average of all members. Third, a target distribution of the members to these roles should be defined. Fourth, the current distribution of roles can be calculated and monitored. Fifth, actions can be defined to move the community nearer to the target distribution.
Apart from this, the automatic identification of roles within a virtual community can be used to bind functionality of the software system supporting the community (e.g. the discussion board software) to it: Identifying a user as visitor can be linked to a guided tour through the virtual community. A novice can be offered friendship or a mentor from the group of leaders. A leader can be offered special functionality like moderating discussions or administrating videos and pictures. And trolls can be limited in their possibility of disturbing the community, for example by disabling their possibility to contribute to discussions.

\section*{\uppercase{Acknowledgments}}

The research leading to these results has received funding from the 
European Community's 7th Framework Program FP7/2007-2013 under 
grant agreement n$^\circ$215453 -- WeKnowIt.

\bibliographystyle{apalike}
\bibliography{cmlm}

\bibliographystyle{IEEEtran}
\bibliography{cmlm}
%



%

\begin{IEEEbiography}{Michael Shell}
Biography text here.
\end{IEEEbiography}

\begin{IEEEbiographynophoto}{John Doe}
Biography text here.
\end{IEEEbiographynophoto}


\begin{IEEEbiographynophoto}{Jane Doe}
Biography text here.
\end{IEEEbiographynophoto}




\end{document}